\documentclass{aastex62}

\submitjournal{ApJ}

\begin{document}

\title{Gravitational Lens System PS J0147+4630 (Andromeda's Parachute): Main Lensing Galaxy 
and Optical Variability of the Quasar Images}
\shorttitle{Gravitational Lens System PS J0147+4630}

\correspondingauthor{Luis J. Goicoechea}
\email{goicol@unican.es, vshal@ukr.net}

\author{Luis J. Goicoechea}
\affil{Departamento de F\'\i sica Moderna \\
Universidad de Cantabria \\
Avda. de Los Castros s/n, 39005 Santander, Spain}

\author{Vyacheslav N. Shalyapin}
\affil{Departamento de F\'\i sica Moderna \\
Universidad de Cantabria \\
Avda. de Los Castros s/n, 39005 Santander, Spain}
\affil{O.Ya. Usikov Institute for Radiophysics and Electronics \\
National Academy of Sciences of Ukraine \\
12 Acad. Proscury St., 61085 Kharkiv, Ukraine}
\affil{Institute of Astronomy of V.N. Karazin Kharkiv National University \\
Svobody Sq. 4, 61022 Kharkiv, Ukraine}



\begin{abstract}
Because follow--up observations of quadruple gravitational lens systems are of 
extraordinary importance for astrophysics and cosmology, we present single--epoch optical 
spectra and $r$--band light curves of PS J0147+4630. This recently discovered system mainly 
consists of four images ABCD of a background quasar around a foreground galaxy G that acts 
as a gravitational lens. First, we use long--slit spectroscopic data in the Gemini 
Observatory Archive and a multi--component fitting to accurately resolve the spectra of A, 
D, and G. The spectral profile of G resembles that of an early--type galaxy at a redshift 
of 0.678 $\pm$ 0.001, which is about 20\% higher than the previous estimate. Additionally, 
the stellar velocity dispersion is measured to $\sim$5\% precision. Second, our early 
$r$--band monitoring with the Liverpool Telescope leads to accurate light curves of the 
four quasar images. Adopting time delays predicted by the lens model, the new lens 
redshift, and a standard cosmology, we report the detection of microlensing variations in C 
and D as large as $\sim$0.1 mag on timescales of a few hundred days. We also estimate an 
actual delay between A and B of a few days (B is leading), which demonstrates the big 
potential of optical monitoring campaigns of PS J0147+4630.
\end{abstract}



\section{Introduction} \label{sec:intro}

Optical light curves of the four images of a quadruple gravitationally lensed quasar (quad)
have a great potential to reveal physical properties of the lensed quasar, the main lensing 
galaxy, and the Universe as a whole. Thus, monitoring campaigns of \object{QSO 2237+0305} 
\citep[Einstein Cross;][]{2000ApJ...529...88W,2002ApJ...572..729A,2008A&A...480..647E} have 
proved to be invaluable tools for unveiling the structure of the quasar accretion disk and 
the composition of the lensing galaxy \citep[e.g.,][]{2002ApJ...579..127S,
2004ApJ...605...58K,2008A&A...490..933E}. Most recently, light curves of other quads also 
led to important astrophysical results, e.g., an estimate of the accretion disk size in 
\object{HE 0435$-$1223} and \object{WFI 2033$-$4723} using microlensing--induced variations 
\citep{2018ApJ...869..132F,2018ApJ...869..106M}, and a robust measurement of the Hubble 
constant $H_0$ from the time delays of \object{HE 0435$-$1223}, \object{RXJ1131$-$1231}, 
and \object{B1608+656} \citep[assuming a flat $\Lambda$CDM cosmology with free energy 
density $\Omega_{\Lambda}$;][]{2017MNRAS.465.4914B}. For a given quad, in addition to 
accurate time delays, the redshift and stellar velocity dispersion of the main lensing 
galaxy (and the properties of the lens environment) are required to obtain strong 
constraints on cosmological parameters \citep[e.g.,][]{2017MNRAS.468.2590S}.  

\citet{2017ApJ...844...90B} reported the serendipitous discovery of the bright quad 
\object{PS J0147+4630} using multiband frames from the Panoramic Survey Telescope and Rapid 
Response System \citep[Pan-STARRS;][]{2016arXiv161205560C}. The four images of this first 
Pan-STARRS lensed quasar are not arranged like a cross, but rather resembling the shape of 
an open parachute (Andromeda's Parachute). An arc--like structure contains the three 
brightest images (A, B and C; $r \sim$ 16$-$16.5 mag), while the fourth component D ($r 
\sim$ 18 mag) is located about 3\arcsec\ from such arc. The main lensing galaxy (G; $r 
\sim$ 20 mag) and D are only $\sim$1\arcsec\ apart. Additionally, spectroscopic 
observations indicated the broad absorption line (BAL) nature of the quasar, which has a 
redshift of 2.341 $\pm$ 0.001 \citep{2017A&A...605L...8L} and is absorbed by several 
intervening metal systems \citep{2018ApJ...859..146R}. From long--slit optical spectroscopy 
with GMOS on the 8.1 m Gemini North Telescope (GNT), \citet{2018MNRAS.475.3086L} also found 
a lens redshift $z_{\rm{G}}$ = 0.5716 $\pm$ 0.0004. These redshifts and {\it Hubble Space 
Telescope} ($HST$) imaging of the lens system allowed \citet{2019MNRAS.483.5649S} to 
predict time delays between the quasar images for a flat $\Lambda$CDM cosmology with $H_0$ 
= 70 km s$^{-1}$ Mpc$^{-1}$ and $\Omega_{\Lambda}$ = 0.7 ($\Omega_m$ = 0.3).

The Gravitational LENses and DArk MAtter (GLENDAMA) project is conducting optical 
observations of a sample of ten lensed quasars with bright images ($r <$ 20 mag) at $1 < z 
< 3$ \citep{2018A&A...616A.118G}. This representative sample includes the quad \object{PS 
J0147+4630}, which is being monitored with the 2.0 m Liverpool Telescope (LT) since 2017 
August. In Section \ref{sec:lens}, we reanalyse the GNT--GMOS data of \object{PS 
J0147+4630} to accurately extract spectra for the three close sources within the slit (A, D 
and G). We then focus on the new spectrum of G, measuring its redshift and the stellar 
velocity dispersion. Implications of this data reanalysis are also discussed in Section 
\ref{sec:lens}. In Section \ref{sec:vari}, we present LT $r$--band light curves of the four 
images of \object{PS J0147+4630} spanning two complete observing seasons from 2017 to 2019,
and show the potential of optical monitoring programmes of this quad. We summarize the 
paper in Section \ref{sec:summ}.  

\section{Reanalysis of the GNT--GMOS spectroscopy: main lensing galaxy} \label{sec:lens}

\citet{2018MNRAS.475.3086L} obtained GNT--GMOS spectroscopic data of \object{PS J0147+4630} 
on 2017 September 2, consisting of 4$\times$1200 s long--slit exposures (B600 grating) 
taken under subarcsecond seeing conditions. The 0\farcs5--width slit with a spatial pixel 
scale of 0\farcs1614 was oriented along the line joining A and D (and crossing G). We 
downloaded these publicly available observations\footnote{Gemini Observatory Archive at 
\url{https://archive.gemini.edu}; Program ID: GN-2017B-FT-4} to reanalyse them. Our aim was 
to accurately resolve the individual spectra of the three sources in the crowded region 
through a well--tested state--of--the--art technique (see below). Before doing the 
extraction of spectra for A, D and G, usual data reductions were performed using the Gemini 
IRAF package\footnote{\url{http://www.gemini.edu/node/11823}}. Regarding the wavelength 
range and dispersion, they were 4403$-$7605 \AA\ and 1.02 \AA\ pix$^{-1}$, respectively. We 
also inferred a resolving power of $\sim$2500 from the 2.24 \AA\ width of the 5577 \AA\ 
[O\,{\sc i}] line in the night airglow. 

We extracted the instrumental spectra of A, D and G by fitting three 1D Moffat profiles in 
the spatial direction for each wavelength bin \citep[e.g.,][]{2007A&A...468..885S,
2017ApJ...836...14S}. For an extended source, an 1D point--spread function (PSF) model does 
not describe its total light. Therefore, since G was treated as a point--like object, we 
actually did not derive the overall spectrum of the galaxy. However, the flux ratio $A/G$ 
is $\sim$250, and the very faint galaxy does not play a relevant role when modeling the 
light distribution along the slit. This distribution is dominated by emission from A and D, 
so the critical issue is our ability to accurately account for the flux of both quasar 
images, minimising the contamination of G. We used $HST$ astrometry 
\citep{2019MNRAS.483.5649S} to set the positions of D and G with respect to A. In regard to 
the 1D PSF model, we checked that a skewed Moffat profile \citep{2014A&A...572A..13S} works 
better than the Gaussian or the symmetric Moffat one \citep[see Appendix 
\ref{sec:specexa};][]{1969A&A.....3..455M}. Thus, we 
adopted a skewed Moffat function with $a_6$ = $a_7$ = 0, i.e., only the asymmetry parameter 
$a_5$ was considered \citep[see Eq. 10 of][]{2014A&A...572A..13S}. We also set the Moffat 
power--law index to an optimal value of 2. In a first fit to the multi--wavelength 1D flux 
distribution, the position (centroid) of A, the width and asymmetry of the Moffat function, 
and the amplitudes of the three components were allowed to vary at each wavelength. We then 
fitted A positions and Moffat structure parameters to smooth polynomial functions of the 
observed wavelength, fixing position--structure parameters to their polynomial values and 
leaving only the three amplitudes as free parameters in a second iteration. 

\begin{deluxetable}{ccccccc}[h!]
\tablecaption{GNT--GMOS--B600 spectra of PS J0147+4630ADG.\label{tab:t1}}
\tablenum{1}
\tablewidth{0pt}
\tablehead{
\colhead{$\lambda$\tablenotemark{a}} & 
\colhead{$F_{\lambda}$(A)\tablenotemark{b}} &  
\colhead{$eF_{\lambda}$(A)\tablenotemark{c}} & 
\colhead{$F_{\lambda}$(D)\tablenotemark{b}} &
\colhead{$eF_{\lambda}$(D)\tablenotemark{c}} & 
\colhead{$F_{\lambda}$(G)\tablenotemark{b}} &
\colhead{$eF_{\lambda}$(G)\tablenotemark{c}}   
}
\startdata
4402.935 & 149.510 & 3.341 & 17.923 & 1.661 & 0.817 & 1.320 \\
4403.956 & 149.636 & 3.344 & 17.938 & 1.663 & 0.818 & 1.321 \\
4404.977 & 149.762 & 3.346 & 17.953 & 1.664 & 0.820 & 1.323 \\
4405.998 & 149.889 & 3.349 & 17.968 & 1.666 & 0.822 & 1.324 \\
4407.019 & 150.016 & 3.352 & 17.983 & 1.667 & 0.823 & 1.325 \\
\enddata
\tablenotetext{a}{Observed wavelength in \AA.}
\tablenotetext{b}{Flux in 10$^{-17}$ erg cm$^{-2}$ s$^{-1}$ \AA$^{-1}$.}
\tablenotetext{c}{Flux error in 10$^{-17}$ erg cm$^{-2}$ s$^{-1}$ \AA$^{-1}$.}
\tablecomments{The spectrum of G is not properly calibrated, since fluxes are 
underestimated in a factor $\sim$5. Table 1 is published in its entirety in the 
machine-readable format. A portion is shown here for guidance regarding its form and 
content.}
\end{deluxetable}

\begin{figure}[h!]
\centering
\includegraphics[width=0.7\textwidth]{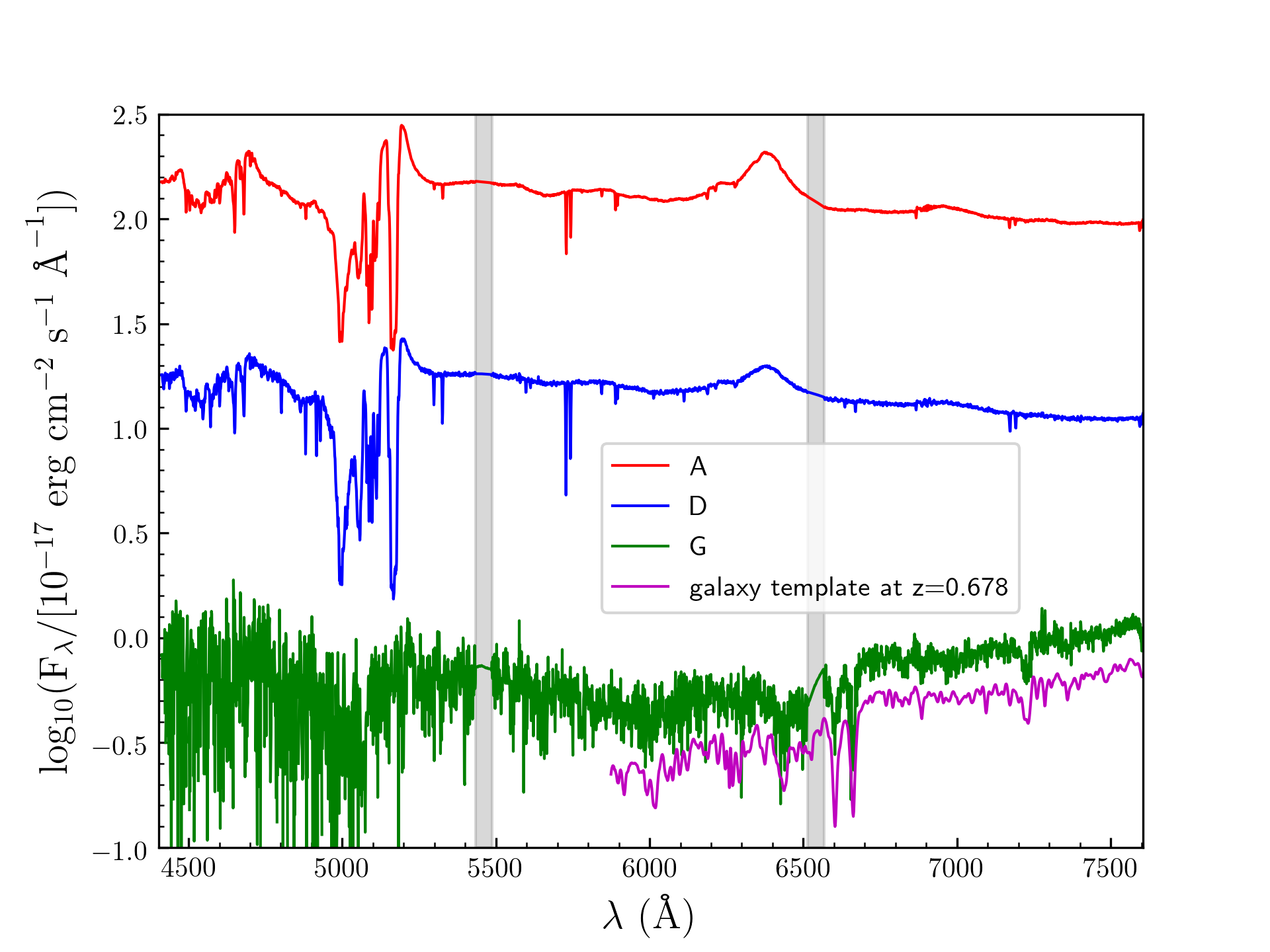}
\caption{GNT--GMOS--B600 spectra of PS J0147+4630ADG in 2017 September. Flux is shown in a 
logarithmic scale to improve visibility, and grey highlighted bars are associated with gaps 
between the detectors. We note that the quasar spectra are reasonably well calibrated in 
flux, while the galaxy spectrum is not. The red--shifted ($z$ = 0.678) template of an 
early--type galaxy is also displayed for comparison purposes (see main text).}
\label{fig:f1} 
\end{figure}

\begin{figure}[h!]
\centering
\includegraphics[width=0.7\textwidth]{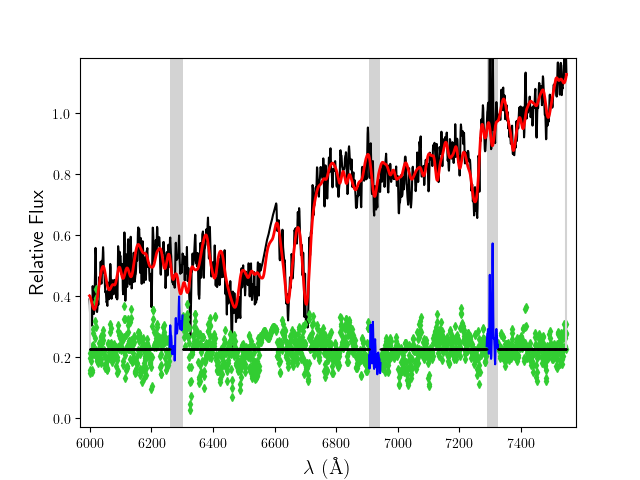}
\caption{Stellar kinematics from the spectrum of the main lensing galaxy. The black line 
describes the observed spectrum, while the red line is the pPXF fit for the stellar 
component. Green symbols show the fit residuals, which are vertically shifted for clarity 
(see the horizontal line). The grey highlighted regions (residuals in blue) were excluded 
during the fitting process.}
\label{fig:f2} 
\end{figure}

The four instrumental spectra of each source (one per exposure) were calibrated in flux and 
combined into a single spectral energy distribution. To carry out the flux calibrations, we 
used GNT--GMOS--B600 0\farcs5--width slit observations of the standard star EG 131 
\citep{1999PASP..111.1426B} on 2017 September 4. Final spectra of A, D and G are included 
in Table~\ref{tab:t1} and plotted in Figure~\ref{fig:f1}. A comprehensive analysys of the 
quasar spectra will be presented in a forthcoming paper. Here, we concentrate on the 
optical spectrum of the main lensing galaxy. The spectral slopes of G, as well as its 
Ca\,{\sc ii} $H\&K$ doublet at about 6600\&6660 \AA\ and its G--band around 7225 \AA, are 
in very good agreement with an early--type galaxy template\footnote{SDSS spectral template 
No. 23 at \url{http://classic.sdss.org/dr7/algorithms/spectemplates/index.html}} at $z$ = 
0.678 (see Figure~\ref{fig:f1}). After doing an initial estimate for $z_{\rm{G}}$ from the 
observed absorption features, we accurately measured the lens redshift using the Penalized 
PiXel--Fitting (pPXF) 
package\footnote{\url{https://www-astro.physics.ox.ac.uk/~mxc/software/}} 
\citep{2004PASP..116..138C,2017MNRAS.466..798C}. This software compares the spectrum of G 
and stellar spectra, providing a correction to the initial value of $z_{\rm{G}}$ and 
measuring the stellar velocity dispersion. From the MILES Library of Stellar 
Spectra\footnote{\url{http://miles.iac.es/}} \citep{2006MNRAS.371..703S,
2011A&A...532A..95F}, we obtained $z_{\rm{G}}$ = 0.678 $\pm$ 0.001 and $\sigma_{\rm{G}}$ = 
313 $\pm$ 14 km s$^{-1}$ (see Figure~\ref{fig:f2}). 

We note that our G spectrum and the redshifted template in Figure~\ref{fig:f1} are very 
similar, while the G spectrum in Fig. 2 of \citet{2018MNRAS.475.3086L} does not resemble 
those of typical galaxies and seems to be heavily contaminated by quasar light. Thus, the 
previous identification of absorption lines of G is unreliable, and we adopt the 
new value of $z_{\rm{G}}$ as the lens redshift. For a flat $\Lambda$CDM cosmology with 
$H_0$ = 70 km s$^{-1}$ Mpc$^{-1}$ and $\Omega_{\Lambda}$ = 0.7, an 18.5\% increase in the 
lens redshift, from 0.572 to 0.678, results in predicted time delays that are $\sim$27\% 
longer than those in Table C1 of \citet{2019MNRAS.483.5649S}. These new delays are $\Delta 
t_{\rm{AB}}$ = $-$2.7, $\Delta t_{\rm{AC}}$ = $-$9, and $\Delta t_{\rm{AD}}$ = 245 days 
($\Delta t_{\rm{XY}} = t_{\rm{Y}} - t_{\rm{X}}$). We also remark that the value of 
$z_{\rm{G}}$ plays a critical role in determining $H_0$ from measured time delays of the  
system. If we assume that $\Omega_{\Lambda} + \Omega_m$ = 1 and 0.6 $\leq \Omega_{\Lambda} 
\leq$ 0.8, then 1.266 $\leq H_0(z_{\rm{G}} = 0.678)/H_0(z_{\rm{G}} = 0.572) \leq$ 1.270.

\section{Optical variability of the quasar images} \label{sec:vari}

\subsection{LT--IO:O light curves} \label{subsec:lt}

\begin{figure}[h!]
\centering
\includegraphics[width=0.7\textwidth]{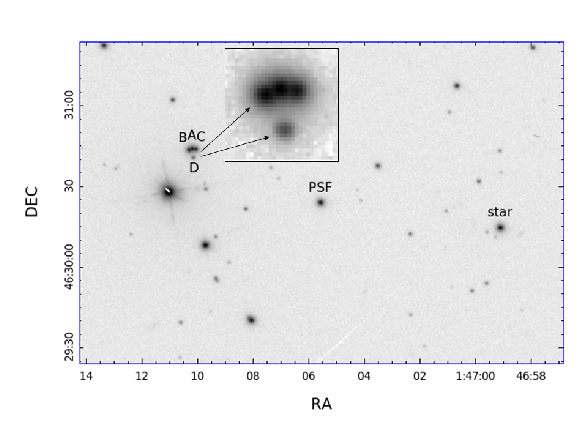}
\caption{LT--IO:O subframe of PS J0147+4630 in the Sloan $r$ band. This subframe of size 
$3\arcmin \times 2\arcmin$ corresponds to the 120 s exposure in subarcsecond seeing on 2018 
August 13. The field of view includes the quad with images ABCD (the strong lensing region 
is zoomed--in into the top sub--panel), the PSF star and the control star (see main text).}
\label{fig:f3} 
\end{figure}

We monitored \object{PS J0147+4630} with the LT from 2017 August to 2018 January and from 
2018 July to 2019 February, i.e., during the first two visibility periods after its 
discovery. All observations were made using the Sloan $r$--band filter on the IO:O optical 
camera (pixel scale of 0\farcs30), and a single 120 s exposure was taken each observing 
night. Although we obtained 84 frames, six of them have poor quality and are not usable for 
doing photometry. Thus, in addition to basic instrumental reductions from the LT--IO:O 
pipeline, we used IRAF software\footnote{\url{https://iraf-community.github.io/}} 
\citep{1986SPIE..627..733T,1993ASPC...52..173T} to remove cosmic rays and bad pixels from 
the remaining 78 frames. The central region of one of these usable frames is shown in 
Figure~\ref{fig:f3}. This includes the four quasar images, as well as an isolated and 
unsaturated star at RA (J2000) = 26\fdg773246 and Dec. (J2000) = +46\fdg506670 ($r$ = 
16.606 mag) that allows us to build up an empirical 2D PSF. A bright field star at RA 
(J2000) = 26\fdg746290 and Dec. (J2000) = +46\fdg504028 is also used as a control object 
having constant brightness $r$ = 15.421 mag (see the middle right side of 
Figure~\ref{fig:f3}).

\begin{deluxetable}{ccccccccc}[h!]
\tablecaption{Photometry of the lens system PS J0147+4630.\label{tab:t2}}
\tablenum{2}
\tablewidth{0pt}
\tablehead{
\colhead{Civil Date\tablenotemark{a}} & 
\colhead{Epoch\tablenotemark{b}} & 
\colhead{FWHM\tablenotemark{c}} & 
\colhead{S/N\tablenotemark{d}} & 
\colhead{A\tablenotemark{e}} & \colhead{B\tablenotemark{e}} & 
\colhead{C\tablenotemark{e}} & \colhead{D\tablenotemark{e}} & 
\colhead{control star\tablenotemark{e}}
}
\startdata
170804 & 7970.051 & 1.06 & 454 & 15.961 & 16.190 & 16.632 & 18.203 & 15.426 \\
170810 & 7976.081 & 0.95 & 334 & 15.953 & 16.198 & 16.622 & 18.211 & 15.421 \\
170816 & 7982.116 & 0.75 & 402 & 15.956 & 16.190 & 16.623 & 18.223 & 15.408 \\
170825 & 7991.048 & 0.95 & 395 & 15.960 & 16.208 & 16.634 & 18.237 & 15.414 \\
170830 & 7996.225 & 1.27 & 466 & 15.955 & 16.191 & 16.637 & 18.237 & 15.416 \\
\enddata
\tablenotetext{a}{yymmdd.}
\tablenotetext{b}{MJD--50000.}
\tablenotetext{c}{FWHM of the seeing disk in $\arcsec$.}
\tablenotetext{d}{S/N for the PSF star within a circle of radius FWHM.}
\tablenotetext{e}{$r$--SDSS magnitude.}
\tablecomments{Table 2 is published in its entirety in the machine-readable format.
A portion is shown here for guidance regarding its form and content.}
\end{deluxetable}

For each of the 78 fames, the brightness of A, B, C, D, and the control star were extracted 
through PSF fitting, using the IMFITFITS software \citep{1998AJ....115.1377M} and the 
PSF star located in the centre of the field of view in Figure~\ref{fig:f3}. 
In the strong lensing region, ABCD and G were modeled as four point--like sources and a de 
Vaucouleurs profile convolved with the empirical 2D PSF, respectively. Our realistic 
overall model also incorporated several $HST$ constraints: positions of B, C, D, and G with 
respect to A, and structure parameters of G \citep{2019MNRAS.483.5649S}. We applied the 
IMFITFITS code to the best frames to obtain the constant galaxy flux, and then to all 
frames (whatever their quality), fixing the galaxy flux and allowing the remaining 
parameters to vary, i.e., the absolute position of A and the four quasar fluxes (see 
Appendix \ref{sec:photexa}). In 
Table~\ref{tab:t2}, we present values of the full--width at half--maximum (FWHM) of the 
seeing disk and the signal--to--noise ratio (S/N) for the PSF star, along with the 
$r$--band magnitudes of the quasar images and the control star. 

\begin{figure}[h!]
\centering
\includegraphics[width=0.7\textwidth]{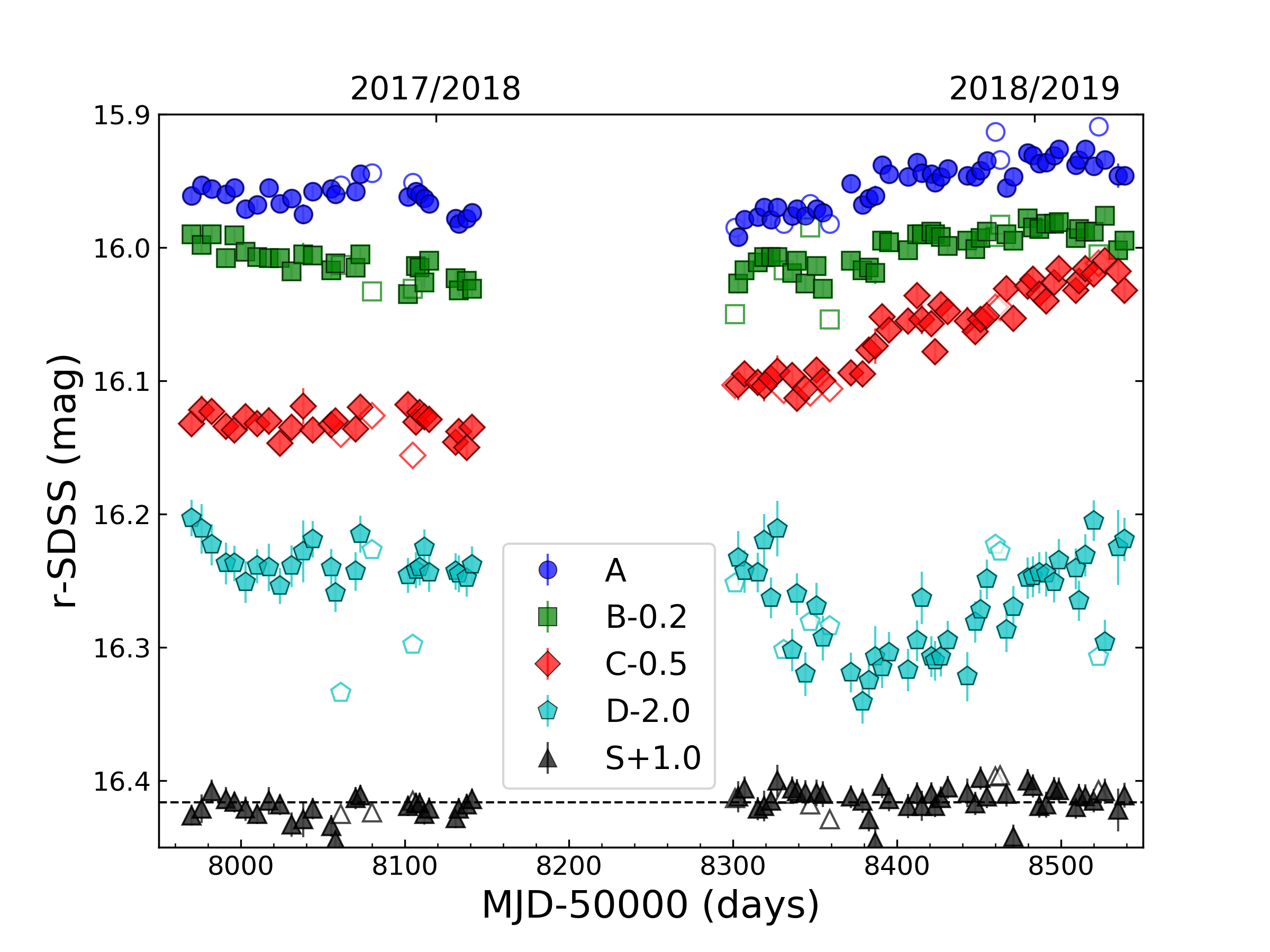}
\caption{LT--IO:O $r$--band light curves of A, B, C, D, and the control star. The curves of 
B, C, D, and S (control star) are shifted by $-$0.2, $-$0.5, $-$2.0, and +1.0 mag, 
respectively, to facilitate comparison. The brightness records from all usable frames (see 
Table~\ref{tab:t2}) are marked with filled and open symbols, while filled symbols trace the 
final light curves after removing 10 epochs (error bars of A, B, and C have sizes 
similar to those of symbols we use; see main text).}
\label{fig:f4} 
\end{figure}

To produce high--quality early light curves of \object{PS J0147+4630}, we selected the 70 
observing epochs in Table~\ref{tab:t2} with FWHM $\leq$ 1\farcs6, and removed two 
additional epochs in which the B magnitude deviates significantly from adjacent values 
(2018 August 16 and 28). In Figure~\ref{fig:f4}, we display our final light curves of the 
quasar images (filled symbols). From the data at these 68 epochs, we also estimated typical 
magnitude errors. For each image, relying on theoretical grounds, we calculated the 
standard deviation between magnitudes having time separations $\leq$ 4.5 days (i.e., 
using 31 pairs of consecutive magnitudes), and then 
divided it by the square root of 2. The resulting errors are 0.0053 (A), 0.0058 (B), 0.0093
(C), and 0.0154 (D) mag. These typical errors were multiplied by the relative S/N at each 
epoch, $\left< \rm{S/N} \right>/(\rm{S/N})_{\rm{epoch}}$, to obtain individual photometric 
uncertainties \citep[$\left< \rm{S/N} \right>$ is the average S/N;][]{2000hccd.book.....H}. 
We achieve $\sim$0.5--1.5\% photometry over a period of about 1.6 years, which includes an
unavoidable visibility gap of more than 5 months. The effective sampling rate (excluding 
the long gap) is $\sim$5 data per month. 

\subsection{What can we learn from early light curves of the lensed quasar?} 
\label{subsec:pot}

\begin{figure}[h!]
	\begin{minipage}[h]{0.5\linewidth}
	\centering
      \includegraphics[angle=-90,width=1.0\textwidth]{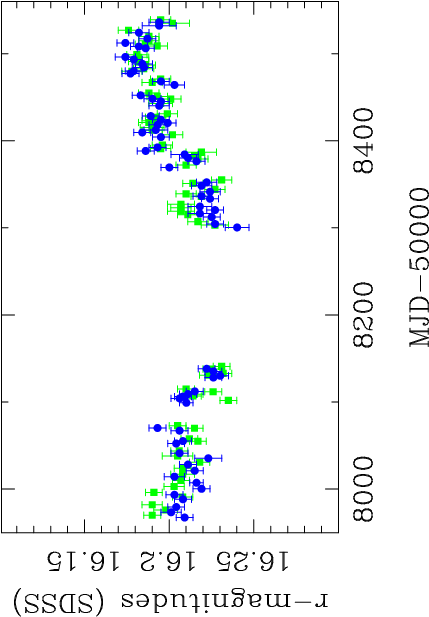}
      \end{minipage}
	\begin{minipage}[h]{0.5\linewidth}
      \centering
      \includegraphics[angle=-90,width=1.0\textwidth]{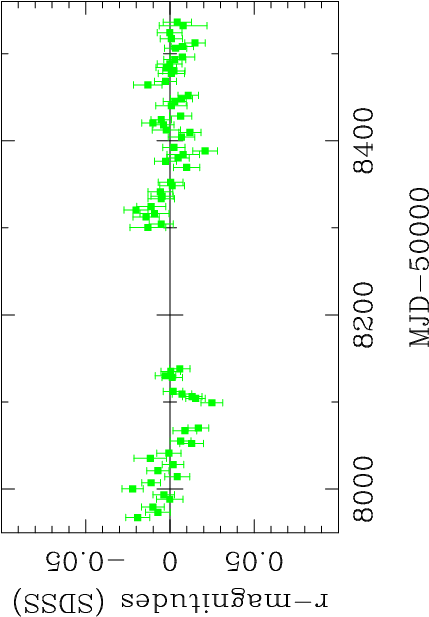}
    	\end{minipage}

\vspace{3.00mm}
	
	\begin{minipage}[h]{0.5\linewidth}
	\centering
      \includegraphics[angle=-90,width=1.0\textwidth]{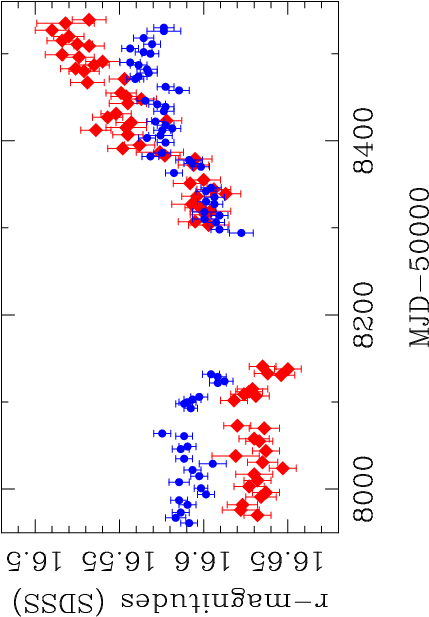}
      \end{minipage}
	\begin{minipage}[h]{0.5\linewidth}
      \centering
      \includegraphics[angle=-90,width=1.0\textwidth]{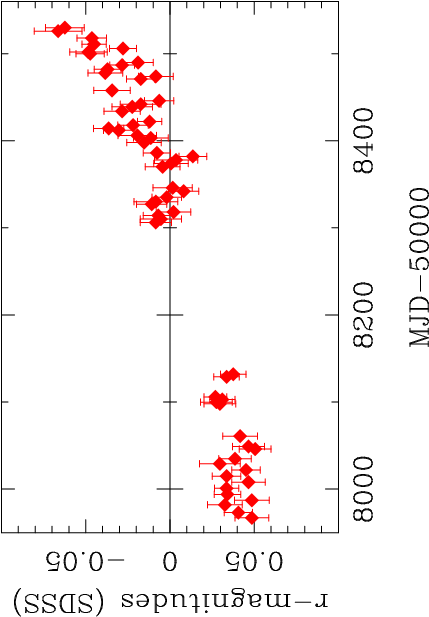}
    	\end{minipage}

\vspace{3.00mm}
	
	\begin{minipage}[h]{0.5\linewidth}
	\centering
      \includegraphics[angle=-90,width=1.0\textwidth]{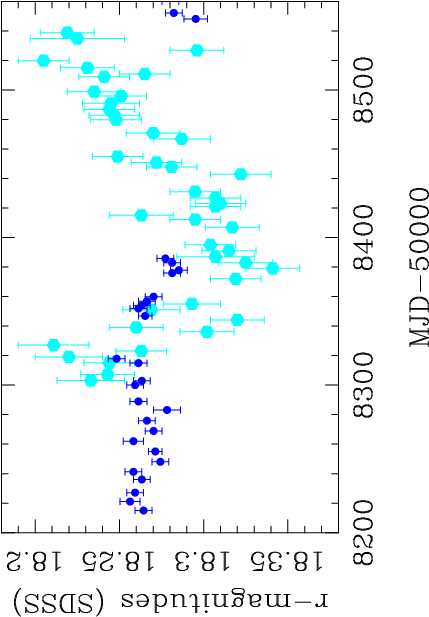}
      \end{minipage}
	\begin{minipage}[h]{0.5\linewidth}
      \centering
      \includegraphics[angle=-90,width=1.0\textwidth]{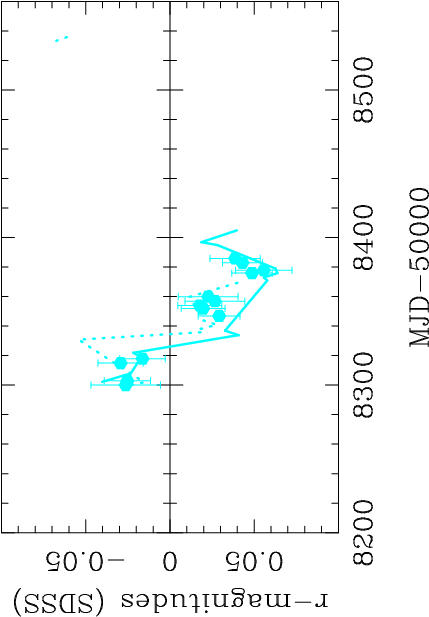}
    	\end{minipage}
\caption{Combined and difference light curves in the $r$ band. The original B, C, and D 
brightness records are compared with magnitude-- and time--shifted light curves of A in the 
top, middle, and bottom panels, respectively. Left: Combined curves. Right: Difference 
curves. In the bottom right panel, in addition to the curve derived through the 
expected delay for $H_0$ = 70 km s$^{-1}$ Mpc$^{-1}$ (filled symbols), we also show 
difference curves for two "extreme" values of $H_0$ \citep[e.g.,][]{2015LRR....18....2J}: 
65 (solid line) and 75 (dotted line) km s$^{-1}$ Mpc$^{-1}$. See main text for details.}
\label{fig:f5} 
\end{figure}

The expected time delays in Section \ref{sec:lens} and the early light curves of ABCD in 
Section \ref{subsec:lt} are useful tools to analyse the origin and properties of the quasar 
variability in the $r$ band. First, we obtained magnitude-- and time--shifted light curves 
of image A as $m_{\rm{AY}}(t) = m_{\rm{A}}(t + \Delta t_{\rm{AY}}) + \left< m_{\rm{Y}} - 
m_{\rm{A}} \right>$ for Y = B, C, and D; and later, each $\left[ m_{\rm{AY}}(t), 
m_{\rm{Y}}(t) \right]$ combined pair was stacked together. These three combined light 
curves are shown in the left panels of Figure~\ref{fig:f5}. Second, we computed difference 
light curves in a standard way, i.e., $m_{\rm{Y}}(t) - m_{\rm{AY}}(t)$ for Y = B, C, and D. 
For a given image Y, the dates in the magnitude-- and time--shifted light curve of A, 
$m_{\rm{AY}}(t)$, were taken as reference epochs. Values of $m_{\rm{AY}}(t)$ were 
subtracted from averaged magnitudes of Y in bins with semiwidth $\alpha$ centred on the 
reference dates. We probed several reasonable values of $\alpha$, obtaining difference 
curves consistent with each other, and then setting $\alpha$ = 5 days (see the right 
panels of Figure~\ref{fig:f5}). The AB comparison (top panels) indicates that both images 
exhibit almost parallel variations, so the difference light curve has a noisy behaviour 
around the zero level. Despite extrinsic (microlensing) variability can be present, it 
should consist of short timescale fluctuations with amplitudes $\leq$ 0.02 mag. However, 
the situation is quite different from the other two AC (middle panels) and AD (bottom 
panels) comparisons. In both cases, we detect significant microlensing variations. The 
difference light curves for C and D include $\sim$0.1 mag fluctuations over timescales from 
100 to 400 days. Some other quads show similar levels of microlensing activity 
\citep[e.g.,][]{2018ApJ...869..132F,2018A&A...616A.118G}. We remark that microlensing
signals in the D image rely on time delays $\Delta t_{\rm{AD}}$ predicted by the Shajib et 
al.'s mass model and plausible values of $H_0$. Despite the true value of $\Delta 
t_{\rm{AD}}$ might be slightly out of the delay range used in this paper, this would not 
produce noticeable changes with respect to the similar signals in the bottom right panel of 
Figure~\ref{fig:f5}.

\begin{figure}[h!]
\centering
\includegraphics[angle=-90,width=0.7\textwidth]{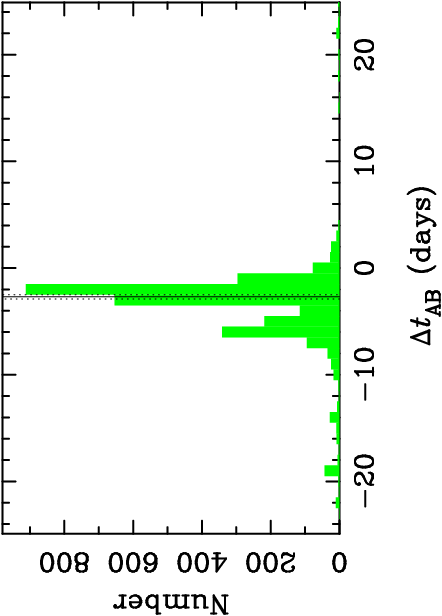}
\caption{Delay histogram from 3000 pairs of simulated curves AB. The best solutions of 
$\Delta t_{\rm{AB}}$ are derived from a $D^2_{4,2}$ estimator ($\delta$ = 10 days) with two
free parameters: time delay and magnitude offset. Predicted time delays for $H_0$ = 
65$-$75 km s$^{-1}$ Mpc$^{-1}$ are shown by a vertical "zipper".}
\label{fig:f6} 
\end{figure}

The LT--IO:O light curves of A and B over the two first monitoring seasons have similar 
shapes, providing evidence for intrinsic variability and enabling us to estimate $\Delta 
t_{\rm{AB}}$. As we only try to show the potential of monitoring campaigns of \object{PS 
J0147+4630}, instead of an exhaustive analysis from several cross--correlation techniques, 
we exclusively used the $D^2_{4,2}$ dispersion estimator \citep{1996A&A...305...97P} to 
match both light curves. This method is simple and very popular, and it works in presence 
of short timescale microlensing \citep[e.g.,][]{2016A&A...596A..77G}. A comprehensive study 
of all time delays will be done when much more extended light curves of the quad are 
available. First, in order to obtain $\Delta t_{\rm{AB}}$ and a constant magnitude offset 
$\Delta m_{\rm{AB}} = m_{\rm{B}}(t) - m_{\rm{A}}(t + \Delta t_{\rm{AB}})$, we focused on a 
biparametric $D^2_{4,2}$. Second, we checked that reasonable values of the decorrelation 
length ($\delta \geq$ 10 days) produce smooth dispersion spectra having minima at $\Delta 
t_{\rm{AB}} <$ 0, and then chose $\delta$ = 10 days. Third, we generated 3000 simulated 
light curves of each quasar image at epochs equal to those of observation. Each observed 
magnitude was modified by adding normally distributed random numbers with zero mean and 
standard deviation equal to the measured uncertainty. Fourth, the $D^2_{4,2}$ estimator was 
applied to all (A, B) pairs of simulated curves. The distribution of magnitude offsets led
to $\Delta m_{\rm{AB}}$ = 0.248 $\pm$ 0.001 mag (1$\sigma$ confidence interval), and the 
time delay histogram is depicted in Figure~\ref{fig:f6}. This yields an 1$\sigma$ interval 
$\Delta t_{\rm{AB}}$ = $-$2.6$^{+1.1}_{-3.2}$ days, indicating that B is leading. 
Additionally, the median delay is in very good agreement with the expected value for 
$H_0$ = 70 km s$^{-1}$ Mpc$^{-1}$ (vertical solid line in Figure~\ref{fig:f6}; see 
the end of Section \ref{sec:lens}). Figure~\ref{fig:f6} also displays a vertical 
"zipper" indicating the narrow range of expected delays for $H_0$ = 65$-$75 km s$^{-1}$ 
Mpc$^{-1}$. 

\section{Summary} \label{sec:summ}

Within the framework of the GLENDAMA project, we are analysing archive data and conducting
follow--up observations of a sample of ten gravitationally lensed quasars 
\citep{2018A&A...616A.118G}, and this paper focuses on the recently discovered Pan--STARRS 
quad \object{PS J0147+4630} \citep{2017ApJ...844...90B}. After retrieving long--slit 
spectroscopic data in the Gemini Observatory Archive \citep{2018MNRAS.475.3086L}, we use a 
robust multi--component fitting \citep[e.g.,][]{2007A&A...468..885S,2017ApJ...836...14S} to
extract individual spectra of the main lensing galaxy and two quasar images in the observed 
crowded region. Despite both quasar spectra contain valuable imprints of intervening 
objects, we only study in detail the optical spectrum of G. This agrees very well with the
spectral profile of an early--type galaxy at a redshift of 0.678, so the previous redshift
determination through a heavily contaminated spectral energy distribution is very likely to 
be biased. An accurate estimate of $z_{\rm{G}}$ is crucial to predict time delays and carry
out cosmological studies \citep[e.g.,][]{2015LRR....18....2J}. We also measure a stellar 
velocity dispersion $\sigma_{\rm{G}}$ = 313 $\pm$ 14 km s$^{-1}$, which is relevant, along 
with other results from ongoing observational efforts \citep[e.g., macrolens flux ratios 
could be revealed by radio imaging with the Very Large Array;][]{2017ApJ...844...90B}, to 
better constrain the lensing mass distribution.

We have also conducted an early $r$--band monitoring of the four quasar images ABCD with 
the Liverpool Telescope, and the corresponding light curves are used to probe the potential 
of optical monitoring campaigns. We take the new redshift $z_{\rm{G}}$ = 0.678 to properly 
modify the time delays predicted by \citet{2019MNRAS.483.5649S}, which permits us to 
construct reliable combined and difference light curves. These curves indicate that C and D 
images are affected by microlensing effects presumably produced by stars within G, and the 
microlensing--induced variations are promising tools to constrain the accretion disk size 
in \object{PS J0147+4630} \citep[e.g.,][]{2018ApJ...869..132F,2018ApJ...869..106M}. 
\citet{2018MNRAS.475.3086L} also pointed out the presence of microlensing when comparing 
his single--epoch Gemini spectra of A and D. In addition, we find that A and B vary in an 
almost parallel way, suggesting that we basically detect intrinsic activity of the source 
quasar in both images. Using a single magnitude offset and a time delay to cross--correlate 
the light curves of A and B, we obtain $\Delta t_{\rm{AB}}$ = $-$2.6$^{+1.1}_{-3.2}$ days.  
Hence, as expected from lens modelling, A arrives a few days later than B. Although the 
$\Delta t_{\rm{AB}}$ estimation might be improved by considering two magnitude offsets or 
other sophisticated approaches, more extended brightness records are required to 
disentangle intrinsic from extrinsic (microlensing) signals in ABCD, and thus, to 
accurately measure the three independent delays $\Delta t_{\rm{AB}}$, $\Delta t_{\rm{AC}}$, 
and $\Delta t_{\rm{AD}}$. These delays are essential pieces for a lens--based cosmology 
\citep[e.g.,][]{2017MNRAS.465.4914B}.  

\acknowledgments

We thank the anonymous referee for helpful comments and suggestions to improve the 
presentation of this paper.
The Liverpool Telescope is operated on the island of La Palma by Liverpool John Moores 
University in the Spanish Observatorio del Roque de los Muchachos of the Instituto de 
Astrofisica de Canarias with financial support from the UK Science and Technology 
Facilities Council. This article is also based on observations obtained at the Gemini 
Observatory (acquired through the Gemini Observatory Archive and processed using the Gemini 
IRAF package), which is operated by the Association of Universities for Research in 
Astronomy, Inc., under a cooperative agreement with the NSF on behalf of the Gemini 
partnership: the National Science Foundation (United States), National Research Council 
(Canada), CONICYT (Chile), Ministerio de Ciencia, Tecnolog\'{i}a e Innovaci\'{o}n 
Productiva (Argentina), Minist\'{e}rio da Ci\^{e}ncia, Tecnologia e Inova\c{c}\~{a}o 
(Brazil), and Korea Astronomy and Space Science Institute (Republic of Korea). We are also 
grateful to the MILES, Pan-STARRS, and SDSS collaborations for doing public databases. This 
research has been conducted in the framework of the Gravitational LENses and DArk MAtter 
(GLENDAMA) project, which is supported by the MINECO/AEI/FEDER-UE grant AYA2017-89815-P and 
the University of Cantabria.

%

\vspace{5mm}
\facilities{Liverpool:2m (IO:O), Gemini:Gillett (GMOS)}


\software{IRAF \citep{1986SPIE..627..733T,1993ASPC...52..173T},
 		pPXF \citep{2004PASP..116..138C,2017MNRAS.466..798C},
		IMFITFITS \citep{1998AJ....115.1377M}
          }



\appendix

\section{Extracting individual spectra of A, D, and G} \label{sec:specexa}

\begin{figure}[h!]
\centering
\includegraphics[width=0.7\textwidth]{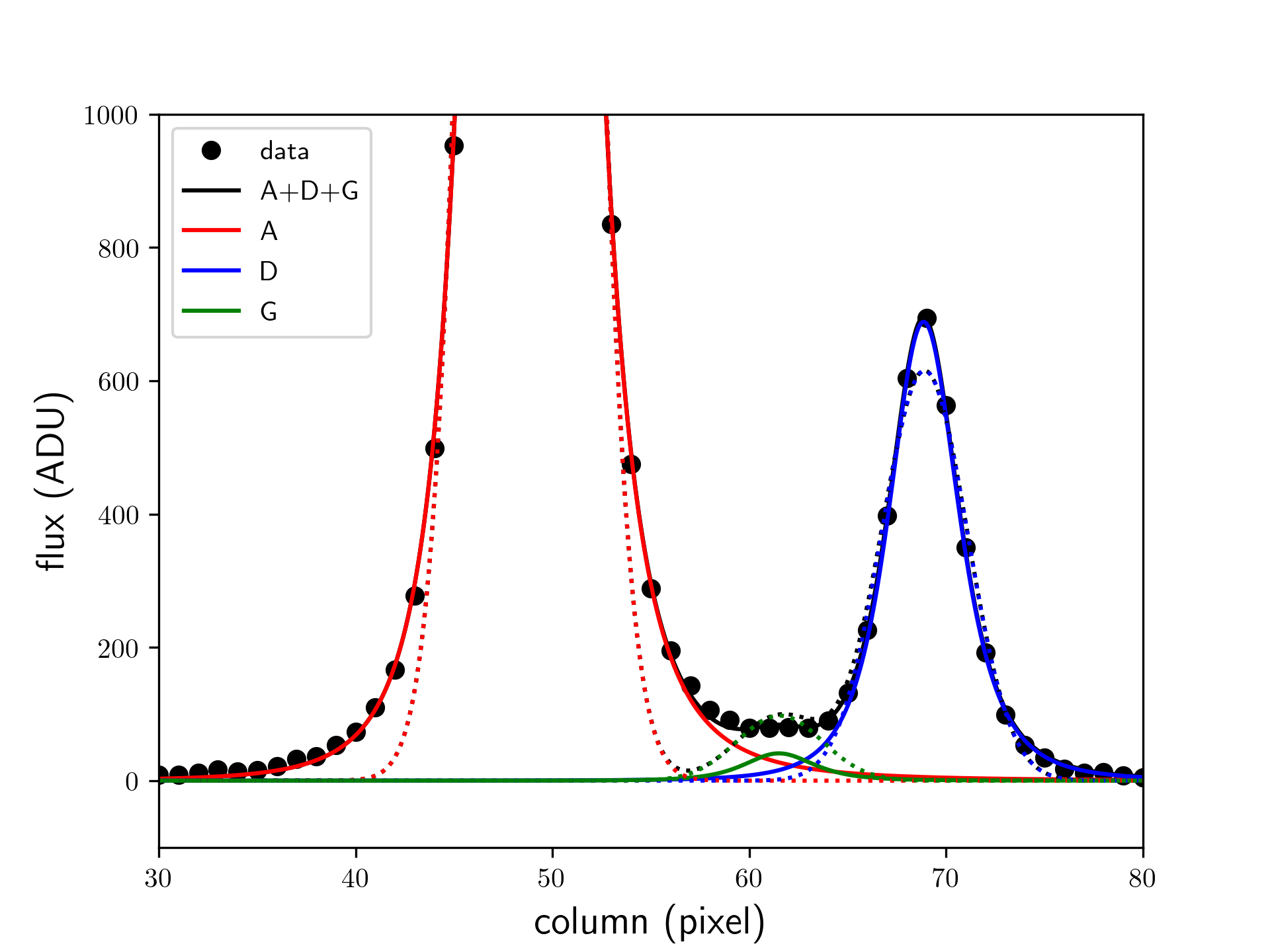}
\caption{1D flux distribution (along the slit) at $\sim$6800 \AA. The black filled circles 
are the measured values, while the solid lines trace the global solution (black), and the 
individual contributions of A (red), D (blue), and G (green) using skewed Moffat profiles.
Using Gaussian profiles, the global solution and the contribution of each source are also 
displayed as dashed lines.}
\label{fig:f7} 
\end{figure}

To illustrate how the multi--component fitting works on instrumental flux distributions 
along the slit, we considered spatial flux distributions at wavelengths around 6800 \AA\ 
for the first 1200s spectroscopic exposure (see Section \ref{sec:lens}). By averaging over 
an 100 \AA\ wavelength interval, we then obtained the spatial profile that appears in 
Figure~\ref{fig:f7} (black filled circles). This 1D flux distribution was fitted to three 
skewed Moffat profiles with the same structure parameters (contributions of the three close 
sources A, D, and G; solid lines with different colors), resulting in a very good global 
solution (see A+D+G in Figure~\ref{fig:f7}). We also show a fit to three Gaussian functions 
(dashed lines) for comparison purposes. These symmetric profiles do not reproduce the data 
in an accurate way, leading to significant fit residuals.

\begin{figure}[h!]
\centering
\includegraphics[width=\textwidth]{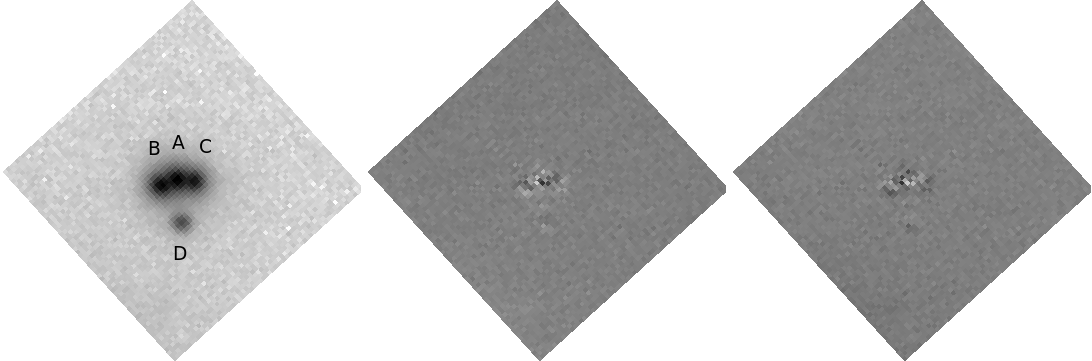}
\caption{Left: 2D flux distribution in the $r$--band on 2018 August 13. Middle: Residual 
fluxes after subtracting only the best--fit model for the quasar images. Right: Residual 
fluxes after subtracting the full best--fit model. Residuals in the middle and right panels 
are linearly scaled between $-$3\% and +3\% of the maximum flux of A. The model consists of 
five components ABCDG and the fitting to data in the left panel is performed with IMFITFITS 
software (see main text).}
\label{fig:f8} 
\end{figure}

\section{Extracting individual $r$--band fluxes of A, B, C, and D} \label{sec:photexa}

As an example of the photometric fitting procedure, we first selected the $r$--band frame 
on 2018 August 13 (see Figure~\ref{fig:f3}). The subframe of interest (containing the lens 
system) is shown in the left panel of Figure~\ref{fig:f8}. Second, the 2D flux distribution 
in this subframe was modelled as four empirical PSFs (quasar images ABCD) plus a de 
Vaucouleurs profile convolved with the empirical PSF (main lensing galaxy G). The best--fit 
model from the IMFITFITS software package \citep{1998AJ....115.1377M} has a reduced 
$\chi^2$ of 1.28, which is a typical value when fitting system subframes at other epochs. 
The middle panel of Figure~\ref{fig:f8} displays the residual signal after subtracting the 
four quasar images. Regarding this panel, it is worth noting that the fluxes of ABCD are 
one or two orders of magnitude larger than the flux of G, and the galaxy is an extended 
source. Additionally, residuals correspond to a relatively short exposure with a 2m 
telescope. Hence, the light distribution in the middle panel is not dominated by emission 
from G. In the right panel of Figure~\ref{fig:f8}, we also show the residual light after 
subtracting the full best--fit model (all sources). Pixels in this last subframe only 
contain the expected noisy signal.

It is worthy to mention that we analyse a crowded system using a well--tested 
photometric method. This worked quite well in other gravitational lens systems, since light 
curves from the IMFITFITS method basically agreed with those from alternative techniques, 
even using different telescopes \citep{2002ApJ...572..729A,2006A&A...452...25U,
2013ApJ...774...69H,2018A&A...616A.118G}. We also note that cross--talk between two close 
sources depends on the seeing, and sometimes it produces evident deviations (with respect 
to adjacent magnitudes) in the light curves of both sources \citep[e.g., see discussion on 
systematic effects and the corresponding outliers in Sect. 4 of][]{2016A&A...596A..77G}. 
However, after removing data from frames with relatively poor seeing and two additional 
epochs producing evident outliers in the brightness record of the B image (see Section 
\ref{subsec:lt}), our final light curves of \object{PS J0147+4630} are not affected by 
significant systematic effects.




\begin{thebibliography}{}

\bibitem[Alcalde et al.(2002)]{2002ApJ...572..729A}
Alcalde, D., Mediavilla, E., Moreau, O., et al.\ 2002, \apj, 572, 729
\bibitem[Berghea et al.(2017)]{2017ApJ...844...90B} 
Berghea, C.~T., Nelson, G.~J., Rusu, C.~E., Keeton, C.~R., \& Dudik, R.~P.\ 2017, \apj, 844, A90
\bibitem[Bessell(1999)]{1999PASP..111.1426B} 
Bessell, M.~S.\ 1999, \pasp, 111, 1426
\bibitem[Bonvin et al.(2017)]{2017MNRAS.465.4914B} 
Bonvin, V., Courbin, F., Suyu, S.~H., et al.\ 2017, \mnras, 465, 4914 
\bibitem[Cappellari(2017)]{2017MNRAS.466..798C}
Cappellari, M.\ 2017, \mnras, 466, 798
\bibitem[Cappellari \& Emsellem(2004)]{2004PASP..116..138C}
Cappellari, M., \& Emsellem, E.\ 2004, \pasp, 116, 138
\bibitem[Chambers et al.(2016)]{2016arXiv161205560C} 
Chambers, K.~C., Magnier, E.~A., Metcalfe, N., et al.\ 2016, eprint arXiv:1612.05560 
\bibitem[Eigenbrod et al.(2008a)]{2008A&A...480..647E} 
Eigenbrod, A., Courbin, F., Sluse, D., Meylan, G., \& Agol, E.\ 2008a, \aap, 480, 647 
\bibitem[Eigenbrod et al.(2008b)]{2008A&A...490..933E} 
Eigenbrod, A., Courbin, F., Meylan, G., et al.\ 2008b, \aap, 490, 933 
\bibitem[Falc\'on-Barroso et al.(2011)]{2011A&A...532A..95F} 
Falc\'on-Barroso, J., S\'anchez-Bl\'azquez, P., Vazdekis, A., et al.\ 2011, \aap, 532, A95 
\bibitem[Fian et al.(2018)]{2018ApJ...869..132F} 
Fian, C., Mediavilla, E., Jim\'enez-Vicente, J., Mu{\~n}oz, J.~A., \& Hanslmeier, A.\ 2018, \apj, 869, A132
\bibitem[Gil-Merino et al.(2018)]{2018A&A...616A.118G}
Gil-Merino, R., Goicoechea, L.~J., Shalyapin, V.~N., \& Oscoz, A.\ 2018, \aap, 616, A118
\bibitem[Goicoechea \& Shalyapin(2016)]{2016A&A...596A..77G}
Goicoechea, L.~J., \& Shalyapin, V.~N.\ 2016, \aap, 596, A77 
\bibitem[Hainline et al.(2013)]{2013ApJ...774...69H} 
Hainline, L.~J., Morgan, C.~W., MacLeod, C.~L., et al.\ 2013, \apj, 774, 69
\bibitem[Howell(2006)]{2000hccd.book.....H} 
Howell, S.~B.\ 2006, Handbook of CCD Astronomy (Cambridge, Cambridge Univ. Press)
\bibitem[Jackson(2015)]{2015LRR....18....2J} 
Jackson, N.\ 2015, Living Rev. Relat., 18, 2 
\bibitem[Kochanek(2004)]{2004ApJ...605...58K} 
Kochanek, C.~S.\ 2004, \apj, 605, 58 
\bibitem[Lee(2017)]{2017A&A...605L...8L} 
Lee, C.-H.\ 2017, \aap, 605, L8
\bibitem[Lee(2018)]{2018MNRAS.475.3086L} 
Lee, C.-H.\ 2018, \mnras, 475, 3086
\bibitem[McLeod et al.(1998)]{1998AJ....115.1377M} McLeod, B.~A., Bernstein, G.~M., Rieke, 
M.~J., \& Weedman, D.~W.\ 1998, \aj, 115, 1377
\bibitem[Moffat(1969)]{1969A&A.....3..455M} 
Moffat, A.~F.~J.\ 1969, \aap, 3, 455 
\bibitem[Morgan et al.(2018)]{2018ApJ...869..106M} 
Morgan, C.~W., Hyer, G.~E., Bonvin, V., et al.\ 2018, \apj, 869, A106 
\bibitem[Pelt et al.(1996)]{1996A&A...305...97P} 
Pelt, J., Kayser, R., Refsdal, S., \& Schramm, T.\ 1996, \aap, 305, 97
\bibitem[Rubin et al.(2018)]{2018ApJ...859..146R} 
Rubin, K.~H.~R., O'Meara, J.~M., Cooksey, K.~L., et al.\ 2018, \apj, 859, A146 
\bibitem[S\'anchez-Bl\'azquez et al.(2006)]{2006MNRAS.371..703S} 
S\'anchez-Bl\'azquez, P., Peletier, R.~F., Jim\'enez-Vicente, J., et al.\ 2006, \mnras, 371, 703 
\bibitem[Sch\"{o}nebeck et al.(2014)]{2014A&A...572A..13S} 
Sch\"{o}nebeck, F., Puzia, T.~H., Pasquali, A., et al.\ 2014, \aap, 572, A13 
\bibitem[Shajib et al.(2019)]{2019MNRAS.483.5649S} 
Shajib, A.~J., Birrer, S., Treu, T., et al.\ 2019, \mnras, 483, 5649 
\bibitem[Shalyapin \& Goicoechea(2017)]{2017ApJ...836...14S}
Shalyapin, V.~N., \& Goicoechea, L.~J.\ 2017, \apj, 836, A14 
\bibitem[Shalyapin et al.(2002)]{2002ApJ...579..127S} 
Shalyapin, V. N., Goicoechea, L.J., Alcalde, D., et al.\ 2002, \apj, 579, 127
\bibitem[Sluse et al.(2007)]{2007A&A...468..885S} 
Sluse, D., Claeskens, J.~F., Hutsem\'ekers, D., \& Surdej, J.\ 2007, \aap, 468, 885
\bibitem[Suyu et al.(2017)]{2017MNRAS.468.2590S} 
Suyu, S.~H., Bonvin, V., Courbin, F., et al.\ 2017, \mnras, 468, 2590
\bibitem[Tody(1986)]{1986SPIE..627..733T} Tody, D.\ 1986, Proc. SPIE, 627, 733 
\bibitem[Tody(1993)]{1993ASPC...52..173T} Tody, D.\ 1993, in ASP Conf. Ser. 52, 
Astronomical Data Analysis Software and Systems II, ed. R.~J. Hanisch, R.~J.~V. 
Brissenden, \& Jeannette Barnes (San Francisco: ASP), 173
\bibitem[Ull\'an et al.(2006)]{2006A&A...452...25U} 
Ull\'an, A., Goicoechea, L.~J., Zheleznyak, A.~P., et al.\ 2006, \aap, 452, 25  
\bibitem[Wo\'zniak et al.(2000)]{2000ApJ...529...88W} 
Wo\'zniak, P.~R., Alard, C., Udalski, A., et al.\ 2000, \apj, 529, 88 

\end{thebibliography}
\end{document}